\documentclass[english,prx, reprint,longbibliography]{revtex4-2}
\usepackage[T1]{
fontenc}
\usepackage[latin9]{inputenc}
\usepackage{xcolor}
\usepackage[unicode=true,pdfusetitle,
bookmarks=true,bookmarksnumbered=false,bookmarksopen=false, breaklinks=true, ,backref=false,colorlinks=true]
 {hyperref}
\hypersetup{citecolor=blue}
\usepackage{babel}
\usepackage{amsmath}
\usepackage{graphicx}

\usepackage{newfloat}
\DeclareFloatingEnvironment[name={Supplementary Figure}]{suppfigure}

\begin{document}

\title{Maximum Caliber Infers Effective Coupling and Response from Spiking Networks}
\author{Kevin S. Chen}
\affiliation{
Princeton Neuroscience Institute, Princeton University, NJ, USA}
\author{Ying-Jen Yang}
\affiliation{Laufer Center of Physical and Quantitative Biology, Stony Brook University, NY, USA}

\date{\today}
\begin{abstract}

The characterization of network and biophysical properties from neural spiking activity is an important goal in neuroscience. 
A framework that provides unbiased inference on causal synaptic interaction and single neural properties has been missing. 
Here we applied the stochastic dynamics extension of Maximum Entropy --- the Maximum Caliber Principle --- to infer the transition rates of network states. Effective synaptic coupling strength and neuronal response functions for various network motifs can then be computed. 
The inferred minimal model also enables leading-order reconstruction of inter-spike interval distribution. 
Our method is tested with numerical simulated spiking networks and applied to data from salamander retina.

\end{abstract}

\maketitle

\section{Introduction}
A network of neurons generates binary time series commonly known as \textit{spike trains} \cite{rieke1999spikes}. The spatiotemporal statistical properties of the spike trains are determined by the synaptic coupling between neurons and by how neurons response to stimuli they received from other neurons and/or from the environment \cite{gerstner_neuronal_2014, koch2004biophysics}. Thus, spike trains contain information about the underlying neuronal network, and can be used to infer network properties of interest \cite{pillow2008spatio, roudi2009statistical}. However, how much information about the network is encoded in simple spiking statistics, and what is the minimal dynamical model based on it?\\

The connection between statistical observables and network properties can be established through building the minimal, unbiased model based solely on the observables. Such minimal model is derived by the Maximum Entropy Principle (Max Ent) as a generic model inference principle \cite{jaynes_information_1957,shore_axiomatic_1980,van_campenhout_maximum_1981}, or its dynamics generalization known as the Maximum Caliber Principle (Max Cal) \cite{jaynes_minimum_1980,csiszar_conditional_1987,presse_principles_2013,chetrite_variational_2015,yang_statistical_2023}. For example, Schneidman \textit{et al.} have shown that the Max Ent model built from the same-time pairwise correlations between neurons can capture the distribution of network spike configurations  \cite{schneidman_weak_2006}; Mora \textit{et al.} have demonstrated that the Max Cal model generated from the autocorrelation of total spike numbers in neighboring time windows can reproduce the size and duration of the spiking-activity avalanches \cite{mora_dynamical_2015}. In a nutshell, Max Ent or Max Cal transforms data to minimal models, from which network properties can be extrapolated and compared to other properties of the network, revealing relations not known before. \\

Here, we are interested in the Max Cal inference of elemental network properties ---primarily the synaptic coupling strength and also the neuronal response functions as well as the inter-spike interval distribution.
A part of this goal has been explored by a recent work from Weistuch \textit{et al.} \cite{weistuch_inferring_2020}. There, correlations between the state variables of neurons \textit{at different times} were used in Max Cal to infer kernel functions promised to reveal the underlying synaptic coupling strength. Approximations for handling the high dimensionality of networks were studied, but the inferred kernels were \textit{not} compared to given underlying coupling strengths. The connections between these kernels and the underlying synaptic coupling strengths are thus unclear.  \\

In the present work, we utilized a more fundamental set of statistical observables to infer effective synaptic coupling and neuronal responses with Max Cal.  
Our method is simple, and yet it correctly identifies the direction and sign of synaptic coupling, as well as the heterogeneity in neuronal responses.
We also showed that the minimal dynamic model from Max Cal can capture the inter-spike interval distributions to the leading order.
Inference performance for synaptic coupling is tested across different neural motifs and subsets of neurons in a network. With such validation from simulated spike trains from biophysical models, we applied our method to salamander retina data \cite{palmer2015predictive, lynn_decomposing_2022} to identify motifs. 

\section{Method \label{sec: theory and method}}

\subsection{Effective Couplings and Neuronal Responses}

\begin{figure}
\begin{centering}
\includegraphics[width=1\columnwidth]{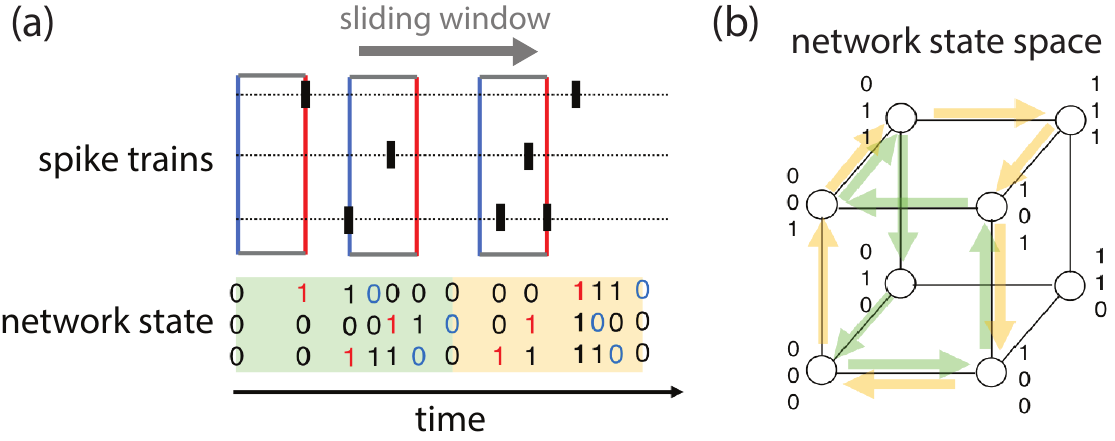}
\par\end{centering}
\caption{\textbf{(a)} Schematic of spike trains converted to a jump process through a sliding window. The spikes enter the window (red) and leave (blue), generating binary events that are converted into the network state show below. \textbf{(b)} Transition of network states shown in a hypercube. The arrows show observed state transitions corresponding to example in (a).  \label{fig: spike to network}}
\end{figure}

A network of $N$ spiking neurons with binary states can have $2^N$ possible configurations ---we call them \textit{network states}. Each time a neuron fires, it triggers its synapses to release neurotransmitters to excite or inhibit its down stream neurons. Such increment of transmitters release lasts at a relevant level in the order of 10 ms \cite{gerstner_neuronal_2014, koch2004biophysics}, and on the downstream neuron end, it takes also in the order of 10 ms for the neuron to reset and forget about the received transmitters. This tells us that the information of a neuron firing can last in the order of 10 ms \cite{gerstner_neuronal_2014}. Within the roughly 10 ms time window when firing information is retained, the downstream neurons have increased/decreased probability of firing. These biophysical arguments are consistent to the empirical observation that time-shifted cross correlation of retinal spikes shows significant decrease in correlation outside of the 20 ms window \cite{schneidman_weak_2006}.\\

Due to the temporary spiking information retainment,  a sliding window of width in the order of 10 ms is used to construct a continuous-time jump process in the network state space from the network spike trains, as shown in Fig. \ref{fig: spike to network}. To our knowledge, this sliding-window construction was first proposed by Lynn et al. 
\cite{lynn_decomposing_2022}. If the neuron has spikes (no spike) in a time window, we assign the neuron to be in its active (silence) state 1 (0). When a spike enters from the right of the window across the red line and if that neuron has not fired within the time window, the network state jumps by flipping that neuron's state from 0 to 1. On the contrary, if a spike leaves the window from the left across the blue line and if that neuron has no other spike within the window, the network state jumps with a 1 to 0 flip on the neuron. With this procedure, the network spike trains of $N$ neurons is converted into a continuous-time jump process with $2^N$ possible network states. \\

A continuous-time process has vanishing probability to have two jumps occur at the same instance \footnote{It happens in real data because spike trains there are discrete-time with finite time binning. In true continuous-time spike trains, the probability is zero}. As each of the $2^N$ has $N$ possible ways to jump out, the network state space is thus a hypercube with $N 2^N$ edges indicating the possible jumps, leading to a \textit{multi-partite} process \cite{horowitz_thermodynamics_2014, wolpert_minimal_2020,lynn_decomposing_2022}. Given long spike trains, we measure the total occupancy time in each network state $\boldsymbol{x}$
\begin{equation}
    \tau_{\boldsymbol{x}} = \text{total time in state }\boldsymbol{x}
\end{equation}
and the counts of transitions from one network state $\boldsymbol{x}$ to another $\boldsymbol{y}$, 
\begin{equation}
    C_{\boldsymbol{x},\boldsymbol{y}} = \text{number of }\boldsymbol{x}\mapsto \boldsymbol{y}.
\end{equation}
From occupancy and transitions, one can compute the empirical distribution $\rho_{\boldsymbol{x}
}$ of each network state $\boldsymbol{x}$
\begin{equation}
    \rho_{\boldsymbol{x}} = \frac{\tau_{\boldsymbol{x}}}{\text{total time } T}
\end{equation}
and the empirical flux $\rho_{\boldsymbol{x},\boldsymbol{y}}$ from state $\boldsymbol{x}$ to $\boldsymbol{y}$
\begin{equation}
    \rho_{\boldsymbol{x},\boldsymbol{y}} = \frac{C_{\boldsymbol{x},\boldsymbol{y}}}{\text{total time } T}.
\end{equation}
Their ratio, in the infinite sampling idealization $T\rightarrow \infty$, gives the tendency rate of making the jump $\boldsymbol{x}\mapsto \boldsymbol{y}$ when the network is in the state $\boldsymbol{x}$:
\begin{equation} \label{eq: rate from C and tau}
    R_{\boldsymbol{x},\boldsymbol{y}} = \lim_{T \rightarrow \infty} \frac{\rho_{\boldsymbol{x},\boldsymbol{y}}}{\rho_{\boldsymbol{x}}} = 
 \lim_{T \rightarrow \infty} \frac{C_{\boldsymbol{x},\boldsymbol{y}}}{\tau_{\boldsymbol{x}}}.
\end{equation}
These rates reveal the synaptic coupling type, the effective strength, as well as the effective neuronal response functions. 
\\

\subsubsection{Effective Synaptic Couplings Cooked Three Ways}
\begin{figure}
\begin{centering}
\includegraphics[width=.75\columnwidth]{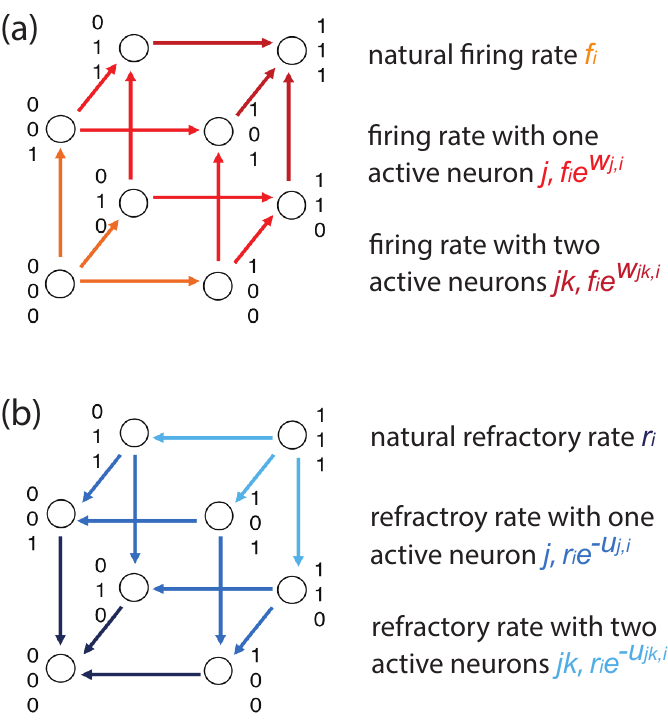}
\par\end{centering}
\caption{\textbf{(a)} Firing rate and coupling parameters governing network state transitions to generate spikes. \textbf{(b)} Refractory rate and coupling parameters governing network state transition to silence. \label{fig: rates}}
\end{figure}

The synaptic coupling information is encoded in the jump rates $R_{\boldsymbol{x},\boldsymbol{y}}$ in multiple aspects. Here we explore them in the simple case of three neurons.  Generalization to $N$ neurons is straightforward.\\

When a network state $\boldsymbol{x}=(000)$ has no firing, then the rate toward the state where only the $i$-th neuron fires --- say the first one $\boldsymbol{y}=(100)$ --- gives a representation of the nature firing rate of the $i$-th neuron: 
\begin{equation} \label{eq: getting f from R}
    f_i = R_{(000),(100)}.
\end{equation}
Comparing this with the rate where the $j$-th neuron, \textit{e.g.} $j=2$, fired before $R_{(010),(110)}$ reveals the effect neuron $j$ causes to neuron $i$. An effective coupling strength can thus be defined by the logarithm of their ratio:
\begin{equation} \label{eq: wji}
    w_{j,i} = \ln \frac{R_{(010),(110)}}{R_{(000),(100)}}
\end{equation}
where the logarithm makes sure the gain (loss) in firing rate ratio on the right-hand side is mapped to a positive (negative) effective coupling strength on the left-hand side. Similarly, the composite effect of two neurons $j$ and $k$ ($k=3$, or more) to neuron $i$ is characterized by the composite effective synaptic input:
\begin{equation} \label{eq: wjki}
    w_{jk,i} = \ln \frac{R_{(011),(111)}}{R_{(000),(100)}}.
\end{equation}
which is the gain in $i$-th neuron firing rate from the previous activation of neurons $j$ and $k$ in the window. \\

With these weights, we can define another effective coupling strengths by 
\begin{equation} \label{eq: w'ji}
    w'_{j,i} = \ln \frac{R_{(011),(111)}}{R_{(001),(101)}} = w_{jk,i}-w_{k,i}.
\end{equation}
This is the effective coupling from $j$ to $i$ when the third neuron $k$ is active, as oppose to $w_{j,i}$ in Eq. \eqref{eq: wji} is when neuron $k$ is silent.
In principle, both of them can provide good coupling inference. Which one of them is more statistical reliable would depend on the data. Here, we note that the two effective couplings $w$ and $w'$ identified by the spiking dynamics in general would not share the same quantitative level but should share the same sign.  A discrepancy in their signs can serve as an indicator for spurious effective couplings, as we will demonstrate with examples later.\\

Synaptic coupling affects not just the firing rates but also the refractory rates, based on which effective couplings could also be defined, as illustrated in Fig. \ref{fig: rates}(b). Since an excitatory coupling will reduce the refractory rate, we could also define the effective coupling by \begin{subequations}
\begin{align}
    u_{j,i} &= -\ln \frac{R_{(110),(010)}}{R_{(100),(000)}} \label{eq: uji}\\
    u_{jk,i} &= -\ln \frac{R_{(111),(011)}}{R_{(100),(000)}}.
\end{align}
\end{subequations}
Biophysically, one would expect that the couplings inferred by $u_{j,i}$ to be less prominent due to neuron's strong self refractory. Nevertheless, later in Fig. \ref{fig: LIF}, we show that these effective coupling strengths are all consistent to each others.


\subsubsection{Effective Neuronal Response Functions}
 By using $\sum_{j(\neq i)}w_{j,i}$ as the effective total ``synaptic current'' received by neuron $i$, denoted as $I_i$, and firing rate $R_i$ as the $i$-th neuron firing rate under different network firing patterns, we can also construct effective response function of neuron $i$, denoted as $R_i=\varphi(I_i)$, based on the network-state jump rates $R_{\boldsymbol{\xi},\boldsymbol{\eta}}$:
\begin{subequations}
\label{eqs: R=varphi(I)}
\begin{align}
     f_i =& \varphi(0) \\
     f_i e^{w_{j,i}} =& \varphi(w_{j,i})\\
     f_i e^{w_{jk,i}} =& \varphi(w_{j,i}+w_{k,i})\\
     \vdots \nonumber
\end{align}
\end{subequations}
in which more points could be obtained from the multi-neuron effects from more neurons $w_{jkl,i}$,$w_{jklm,i}$, ... etc.
The curve $\varphi$ could be effectively understood as the firing-rate to stimuli response function of the neurons, and can be used to determine the heterogeneity among neurons.  An example for three neurons will be shown in Fig. \ref{fig: LIF}(c).\\


\subsection{Maximum Caliber infers Continuous-time Markov chain as the minimal model}

We have shown above that the network state jump rates $R_{\boldsymbol{\xi,\eta}}$ defined from network state occupancy time $\tau$ and transition counts $C$ can be used to reveal effective network properties, such as the coupling strength and the neuronal response functions. How many other network properties could be captured by the information encoded in these rates $R_{\boldsymbol{\xi,\eta}}$? To see this, we use the dynamics extension of the Maximum Entropy principle --- Maximum Caliber principle --- to build the minimal dynamical model based on the observables $\tau$ and $C$, or equivalently $(\rho_{\boldsymbol{\xi}},\rho_{\boldsymbol{\xi,\eta}})$,  that define the rates $R_{\boldsymbol{\xi,\eta}}$.   \\



The Maximum Entropy principle (Max Ent), originated from equilibrium statistical physics, has been generalized as an inference principle to generate an unique, unbiased model with minimal update to a prior model \cite{jaynes_information_1957,shore_axiomatic_1980,van_campenhout_maximum_1981,presse_principles_2013, jizba_maximum_2019,caticha_entropy_2021}. In a nutshell, given a prior probability distribution model $P_0(z)$ and the set of candidate posteriors $\mathcal{P}_c$ satisfying a set of constraints $c$, Max Ent finds the unique posterior $P^*(z)$ that is Kullback-Leibler (KL) closest to the prior $P_0$:
\begin{align} 
    P^* &=\arg \min_{P \in \mathcal{P}_c} D(P||P_0) \nonumber\\
    &= \arg \min_{P \in \mathcal{P}_c} \sum_z P(z) \log \frac{P(z)}{P_0(z)}.
\end{align} 
As explained intuitively by Jaynes \cite{jaynes_information_1957}, since Kullback-Leibler (KL) divergence $D(P||P_0)$ is the unique measure of information gain \cite{shannon_mathematical_1948,khinchin_mathematical_1957}, the Max Ent posterior $P^*$ is thus unbiased in that it is the minimal model with no other information added other than the prior $P_0$ as the reference model and the constraints $c$. 
Jaynes' intuitive explanation has been made mathematically precise recently through the axiomatization of Max Ent as a statistical model inference \cite{shore_axiomatic_1980,presse_nonadditive_2013,tsallis_conceptual_2015,presse_reply_2015,jizba_maximum_2019,caticha_entropy_2021}.
In its application, Max Ent transforms constraints (our knowledge about the system) into a minimal model $P^*$ that can be used to find out what network properties the constraints are enough to predict \cite{schneidman_weak_2006}.
\\

With $z$ taken as the system's state, Max Ent only infers static state distribution model, not models about dynamics. To apply Max Ent to dynamics, Jaynes further proposed taking $z$ as paths $\{x(t),0\le t\le T\}$, and called such Max Ent of dynamics the principle of \textit{Maximum Caliber} (Max Cal) \cite{jaynes_minimum_1980}. For recent reviews, see, \textit{e.g.}, \cite{presse_principles_2013,pachter_entropy_2024}. When considering only the averages of time-additive state and transition observables, $\langle \sum_t F_{x(t)} \rangle$ and $\langle \sum_t G_{x(t-\mathrm{d}t),x(t)} \rangle$, it has been shown that Max Cal produces Markov dynamics as the minimal dynamical model \cite{csiszar_conditional_1987,ge_markov_2012,lee_derivation_2012}. For examples, the counting observables we considered above --- distribution $\rho_{\boldsymbol{\xi}}$ and flux $\rho_{\boldsymbol{\xi},\boldsymbol{\eta}}$--- correspond to choosing delta functions for $F$ and $G$ in continuous time with delta functions \cite{ge_markov_2012, yang_statistical_2023}:
\begin{align}
    \rho_{\boldsymbol{\xi}}[\boldsymbol{x}(t)] &=  \int_0^T \frac{\delta_{\boldsymbol{\xi},\boldsymbol{x}(t)}}{T} \mathrm{d}t\\
     \rho_{\boldsymbol{\xi},\boldsymbol{\eta}}[\boldsymbol{x}(t)] &=  \sum_{t_m} \frac{\delta_{\boldsymbol{\xi},\boldsymbol{x}(t_{m-1})}\delta_{\boldsymbol{\eta},\boldsymbol{x}(t_m)}}{T} 
\end{align}
where $t_m$ are jump times and $\boldsymbol{x}(t_m)$ is the state immediately after the jump. Given these counting observables $\rho_{\boldsymbol{\xi}}$ and $\rho_{\boldsymbol{\xi,\eta}}$, Max Cal produces effective Markov models.
\\

If possible, we would measure these counting observables $\rho_{\boldsymbol{\xi}}$ and $\rho_{\boldsymbol{\xi},\boldsymbol{\eta}}$ for arbitrary long paths ---ideally for infinitely long paths so that they become their asymptotic, steady-state values as probability distribution $\pi_{\boldsymbol{\xi}}$ and flux $p_{\boldsymbol{\xi},\boldsymbol{\eta}}$. 
Max Ent, as a theory, is built for the idealized infinite sampling that gives us the expected values $(\pi,p)$ \cite{yang_statistical_2023}.
When we have finite but large data, we would treat $\rho_{\boldsymbol{\xi}}$ and $\rho_{\boldsymbol{\xi},\boldsymbol{\eta}}$  as surrogates to $\pi_{\boldsymbol{\xi}}$ and $p_{\boldsymbol{\xi},\boldsymbol{\eta}}$, as if the data were infinitely long, an idea from thermodynamics of building a theory for infinitely large system and applying it to finite system as a leading order approximation \cite{anderson_more_1972,yang_statistical_2023}. Thus, the Max Cal of time homogeneous processes minimizes the steady-state KL rate that dominates the KL divergence $D(R||R^0)$ in the long-term limit: 
\begin{subequations}
\begin{align}
    d(R||R^0)&=\lim_{T\rightarrow \infty} \frac{D(R||R^0)}{T} \\  
    =& \sum_{\boldsymbol{\xi}\neq \boldsymbol{\eta}} \pi_{\boldsymbol{\xi}} R_{\boldsymbol{\xi,\eta}} \log \frac{R_{\boldsymbol{\xi,\eta}}}{R^0_{\boldsymbol{\xi,\eta}}} + \pi_{\boldsymbol{\xi}} R_{\boldsymbol{\xi,\eta}} -  \pi_{\boldsymbol{\xi}} R^0_{\boldsymbol{\xi,\eta}} \label{rate in R}
\end{align}
\end{subequations}
where the steady-state distribution $\pi_i(R_{\boldsymbol{\xi,\eta}})$ is a function of the transition rates $R_{ij}$. Given a set of steady-state statistics $(\pi_{\boldsymbol{\xi}},p_{\boldsymbol{\xi},\boldsymbol{\eta}})$ estimated by $(\rho_{\boldsymbol{\xi}},\rho_{\boldsymbol{\xi},\boldsymbol{\eta}})$ from the data, Max Cal finds the continuous-time Markov chain (CTMC) with the posterior transition rates $R^*_{ij}$ that minimizes $d$ and satisfies the constraints:
\begin{equation} \label{eq: Max Cal}
    R^*_{\boldsymbol{\xi,\eta}} = \arg \min_{R\in \mathcal{R}_c} d(R||R^0) 
\end{equation}
where $\mathcal{R}_c$ collects all possible processes $R_{\boldsymbol{\xi,\eta}}$ that satisfies a given set of constraints $c$, \textit{e.g.} a bunch of distribution and flux measurements, no need to be the full set.\\

\begin{figure}
\begin{centering}
\includegraphics[width=1\columnwidth]{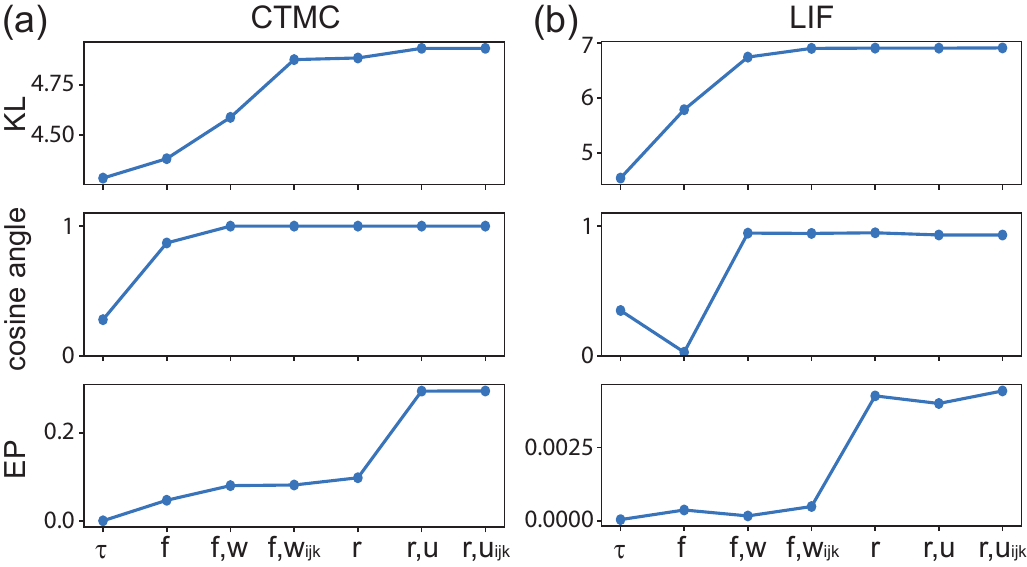}
\par\end{centering}
\caption{{Learning curve with Max Cal inference.} \textbf{(a)} Inference for continuous-time Markov chain (CTMC) across accumulation of different groups of observables in the x-axis. The optimized KL term (top), cosine angle between inferred and ground-truth weights (middle), and the notion of time irreversibility called ``entropy production'' in stochastic thermodynamics (EP, bottom) are shown. \textbf{(b)} Same as (a) but for a finite spike train simulated from a motif of three leaky-integrate-and-fire (LIF) neurons.  \label{fig: dof}}
\end{figure}

These posterior rates $R^*_{\boldsymbol{\xi,\eta}}$ could then give the inferred effective synaptic couplings and neuronal responses based on Eq. \eqref{eq: wji}, \eqref{eq: w'ji},  \eqref{eq: uji}, and \eqref{eqs: R=varphi(I)}. And the posterior KL-rate $d^*=d(R^*||R^0)$ represents the \textit{information gain} from the set of constraints $c$. With more and more constraints feed into Max Cal, we gain more information, and $d^*$ increases monotonically as shown in Fig. \ref{fig: dof}.\\

In Fig. \ref{fig: dof}, we consider accumulating groups of constraints in Max Cal. The first case $\tau$ represents feeding in the full set of state occupancy times $\{\tau_{\boldsymbol{\xi}}\},\boldsymbol{\xi}=\{\text{all network states}\}$, which can give us full information about $\rho_{\boldsymbol{\xi}}$ as estimation to $\pi_{\boldsymbol{\xi}}$. The second point $f$ represents the accumulated set where we measure both $\tau_{\boldsymbol{\xi}}$ and the transition counts $C_{\boldsymbol{\xi,\eta}}$ that can give us the parameters $f$ according to Eq. \eqref{eq: getting f from R}. In the three-neuron example here, these counts are $C_{000,001},C_{000,010},$ and $C_{000,100}$. Similarly for the rest of the plot, we accumulated more and more transition counts to cover degrees of freedom specified by $f_i e^{w_{j,i}}, f_i e^{w_{jk,i}}$, and the refractory rates. \\

For each set of constraints, we computed the Max Cal posterior $R^*_{\boldsymbol{\xi,\eta}}$ and associated with it the posterior KL rate $d^*$. As $d^*$ is a convex function in the space of all models, accumulating more constraints (toward the right of the plots) will lead to bigger posterior $d^*$, representing an accumulation of information. From these posterior rates $R^*_{\boldsymbol{\xi,\eta}}$, the inferred network properties such as the effective coupling strength $w_{j,i}$ can be computed.\\

In the second row of Fig. \ref{fig: dof}, we showed the cosine angle
between the inferred $w_{j,i}$ and the ground truth coupling strength,
\begin{equation}
    \text{cosine angle} = \frac{\boldsymbol{w}_{\rm infer} }{|| \boldsymbol{w}_{\rm infer} || } \cdot \frac{ \boldsymbol{w}_{\rm truth}}{|| \boldsymbol{w}_{\rm truth} || },
\end{equation}
in a CTMC toy network of three neurons with ``data'' computed by analytic calculations (infinite data) and a spiking neuron network with leaky-integrate-and-fire (LIF) neurons generating finite spike trains. The cosine angle saturates as expected since, after the third group $(f,w)$, all information about $w_{j,i}$ has been added into Max Cal and the remaining updates are not about them.\\

In the third row of Fig. \ref{fig: dof}, we computed the effective time irreversibility in the inferred Markov dynamics, also known as the ``entropy production'' (EP) in stochastic thermodynamics \cite{seifert_stochastic_2019,yang_unified_2020,van_den_broeck_ensemble_2015,jarzynski_equalities_2011,peliti_stochastic_2021,lynn_decomposing_2022}:
\begin{equation}
    \text{EP}=\sum_{\boldsymbol{\xi,\eta}} p^*_{\boldsymbol{\xi,\eta}}\log \frac{p^*_{\boldsymbol{\xi,\eta}}}{p^*_{\boldsymbol{\eta,\xi}}}.
\end{equation}
We note that the EP of the first Max Cal posterior model with only the $\tau$ constraints is always zero. This is because our unit-rate prior is time reversible (detailed balanced), and the information in $\tau$ is independent to the irreversibility \footnote{One of us is preparing a manuscript on this.}, leading to no update on the EP. This demonstrates the minimal/unbiased updating of Max Cal. EP starts to get updates after learning the constraints about the transitions. We also remark that the EPs in these Max Cal models with accumulating constraints, unlike the KL rate $d^*$, are not monotonic. EP is not information; path KL rate is.\\

In general, to saturate the KL rate $d^*$ and ensure the posterior $R^*_{\boldsymbol{\xi,\eta}}$ is free from future updates, we will need to cover all degrees of freedom in the rates $R^*_{\boldsymbol{\xi,\eta}}$. While there are degeneracy in the empirical distributions $\rho_{\boldsymbol{\xi}}$ and fluxes $\rho_{\boldsymbol{\xi,\eta}}$ due to normalization and stationarity, it is convenient to simply put all of them into Max Cal. Note that when including the full set of $(\pi_{\boldsymbol{\xi}},p_{\boldsymbol{\xi,\eta}})$, the Max Cal-inferred transition rates $R^*_{\boldsymbol{\xi,\eta}}$ can be directly computed by using Eq. \eqref{eq: rate from C and tau}, without actually performing the optimization computation, as we shall employ from now on. What Max Cal tells us is that the Markov dynamics with rates $R^*_{\boldsymbol{\xi},\boldsymbol{\eta}}$ computed by the data is the inferred minimal, unbiased model. Max Cal also brings us these learning curves. The early saturation of them could indicate hidden symmetries in the systems and/or relations between the constraints and the network properties of interest. We leave the exploration of this to future studies. 


\section{Results \label{sec: Results}}

\begin{figure}
\begin{centering}
\includegraphics[width=1\columnwidth]{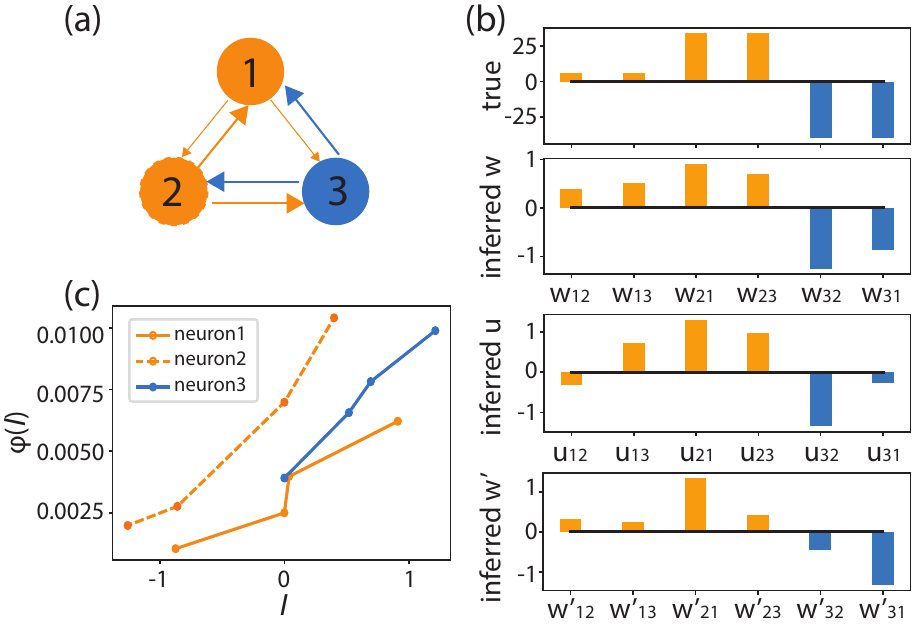}
\par\end{centering}
\caption{Inferring coupling and response in LIF circuits. \textbf{(a)} Three-neuron circuit with balanced E-I. Neuron 1 and 2 are excitatory whereas neuron 3 is inhibitory. \textbf{(b)} The true (top) and inferred weights $w_{ij}$, refractoriness $u_{ij}$, and effective coupling $w'_{ij}$. \textbf{(c)} The inferred response function $\varphi$ for the same three neurons shown in (b). Neuron 2 has a lower firing threshold compared to neuron 1 and 3.
\label{fig: LIF}}
\end{figure}

\subsection{Neuronal Network Model : LIF E-I balanced}

We first tested inference for network coupling and response from spike trains simulated with a 3-neuron circuit (Fig.~\ref{fig: LIF}a).  We modelled a fully connected excitatory-inhibitory (E-I) balanced motif, with two excitatory neurons and an inhibitory neuron that has stronger projection. 
The neurons are leaky-integrate-and-fire (LIF) units that integrate current input and generate discrete spikes when the voltage exceeds a threshold. The synaptic coupling is modelled through a weighted and filtered spike impulse from the pre-synaptic neuron. 
\\





\paragraph{Inferring Coupling Type and Strength $w$}

The inferred weights $w$, refractory coupling $u$, and effective coupling $w'$ have consistent trends as the true synaptic weights underlying LIF circuit (Fig.~\ref{fig: LIF}b). 
Particularly, $w$ correctly reflects the underlying sign and relative amplitude of synaptic connection.
While the computed $w'$ is with larger variations, potentially due to the limited occurrence for the state transitions between more synchronized states such as (011) to (111), the excitation and inhibition inference are nevertheless consistent with the true network.\\ 

To test the robustness of the inference, we explored effects of the time window with a perturbative approach and confirm that the inference is consistent within a range of time window (SFig.~\ref{SI:window}). In addition, we also scanned different noise strength and network weights in the LIF model to characterize effects of signal-to-noise ratio in weight inference (SFig.~\ref{SI:SNR}).\\
 

\paragraph{Inferring Neuronal Responses}

In the simulation, one of the two excitatory neuron (neuron 2) has a lower spiking threshold (-60 mV, compared to the default value at -50 mV). Such neuronal heterogeneity can be inferred from the response curves computed by Eq. \eqref{eqs: R=varphi(I)} shown in Fig.~\ref{fig: LIF}c. Neuron 2 has a higher response curve that reflects its lower spiking threshold. \\

We note that the LIF unit is a generic biophysical model capable of capturing spiking dynamics in complex experimental measurements \cite{teeter2018generalized, ladenbauer2019inferring}. The fact that the jump-process representation of network spike trains can recover the biophysical parameters not only validates the inference method but also suggests application to biological data, which we will visit in last section.

\subsection{Inferring Different Motifs}
\begin{figure}[!t]
\begin{centering}
\includegraphics[width=1.\columnwidth]{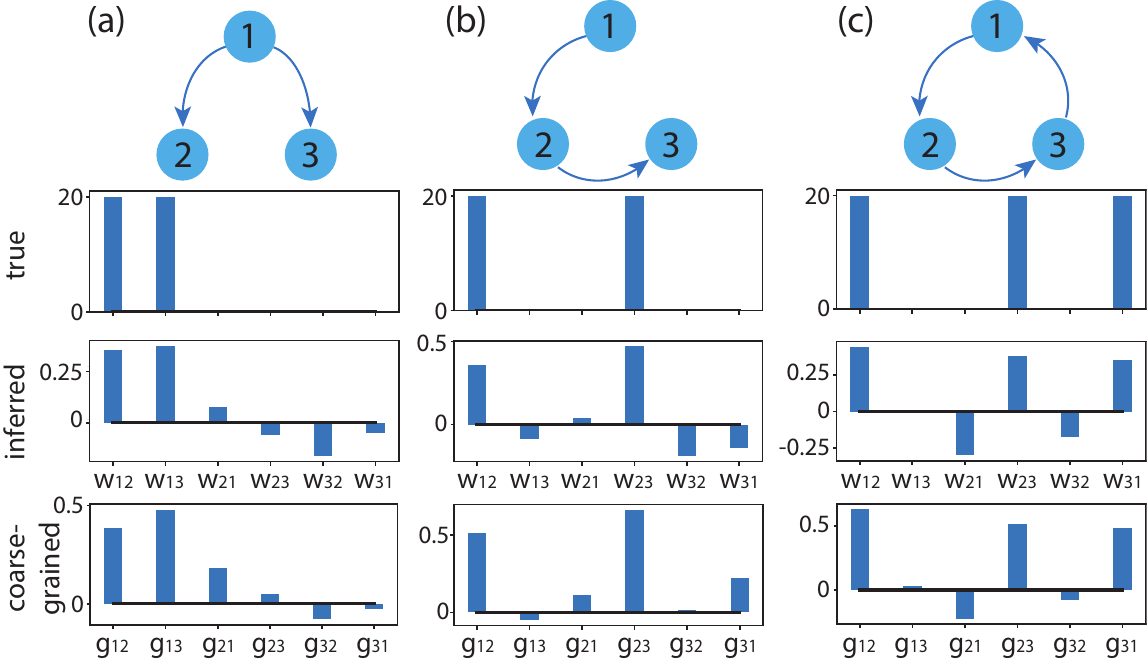}
\par\end{centering}
\caption{Inferring different circuit motifs. \textbf{(a)} Motif with common input shown on top. The true connection and inferred weights $w_{ij}$ shown in the middle. The coupling inferred through coarse-graining $g_{ij}$ is shown on the third row. \textbf{(b)} Same as (a) but for a chain motif. \textbf{(c)} Same as (a) but for a cyclic motif. 
\label{fig: motif}}
\end{figure}

Biological networks are rich in particular motifs to accommodate different functions and information processing \cite{luo2021architectures}. We explored the flexibility of the inference method in identifying different motifs by simulating spike trains from three LIF motifs: 
(1) common input, (2) chain, (3) cyclic motifs (Fig.~\ref{fig: motif}). Across different motifs, we found that the inferred weights correctly identify the true connections.\\

In particular, the common input motif (Fig.~\ref{fig: motif}a) tests for the correlation-causality difference in spike train inference. In this motif, neuron 2 and 3 have highly correlated spiking pattern due to the common and symmetric inputs from neuron 1. 
However, the inferred connectivity correctly identifies that neuron 2 and 3 are not causally connected. The non-connected parameters fluctuate depending on data length and are significantly smaller compared to the true connections. 
 In addition, the chain motif (Fig.~\ref{fig: motif}b) introduces an indirect coupling pathway from neuron 1 to 3 through 2. Our inference method can correctly identify only the direct effect. Similarly, the cyclic motif (Fig.~\ref{fig: motif}c) introduces interconnected spiking correlation. In result, the inferred weights consistently identify the signs of the true connections across all motifs.\\\\

\subsection{Marginalizing Out Unobserved Neurons}

The aforementioned neural motifs can be recovered when spike trains are fully observed. However, in many practical cases, unobserved neurons can have input to the observed ones and influencing the statistics \cite{morrell2021latent}. We explored the effects of unobserved neurons by coarse-graining the 3-neuron analysis and subsampling from a larger network. The validity of this local inference approach allows our method to be applied to larger networks.\\

In the three-neuron cases, in contrast to tracking all three neurons to infer $w_{ij}$, 
we could also take any two of the three neurons and repeat the inference steps for their spiking data alone, ignoring the other neuron. We call this ``coarse-grained'' coupling $g_{ji}$: 
\begin{equation}
    g_{ji} = \log \frac{C_{(...01...),(...11...)}/\tau_{(...01...)}}{C_{(...00...),(...10...)}/\tau
_{(...00...)}}
\end{equation} where $\tau_{(...01...)}$ is the total occupancy time that $\sigma_i=0$ and $\sigma_j=1$ regardless of other neurons, and $C_{(...01...),(...11...)}$ is the total number of times we see transition $01\mapsto 11$ regardless of other neurons.
This is equivalent as marginalizing out other neurons. The coarse-grained coupling $g_{ji}$ are thus related to $w_{j,i}$ and $w_{jk,i}$, which we get from including third neuron: 
\begin{equation}
    g_{ji} = \log \frac{\frac{\tau_{(...010...)}}{\tau_{(...01...)}} e^{w_{j,i}}+\frac{\tau_{(...011...)}}{\tau_{(...01...)}} e^{w_{jk,i}}}{\frac{\tau_{(...000...)}}{\tau_{(...00...)}} +\frac{\tau_{(...001...)}}{\tau_{(...00...)}} e^{w_{k,i}}}
\end{equation}
When coupling from $j$ to $i$ is excitatory, we expect both $w_{j,i}>0$ and $w'_{j,i}>0 \Leftrightarrow w_{jk,i}>w_{k,i}$. These imply $g_{ji}>0$. Therefore, when there is a sign discrepancy between $w_{j,i}$ and $g_{j,i}$, it is an indicator for potential spurious inferred coupling due to finite data. For example, edge 23 in Fig. \ref{fig: motif}(a), and edge 31 in (b).  \\

The coarse-graining analysis only focuses on paired interactions and integrates out the third neuron, but can still show consistent indication of synaptic connections. This is consistent with many past work demonstrating the statistical power of pairwise interactions in the neural network \cite{schneidman_weak_2006, pillow2008spatio}. Motivated by this result, we proceed to explore the effects of unobserved neurons on motif inference in a larger network. \\

\begin{figure}[!t]
\begin{centering}
\includegraphics[width=1\columnwidth]{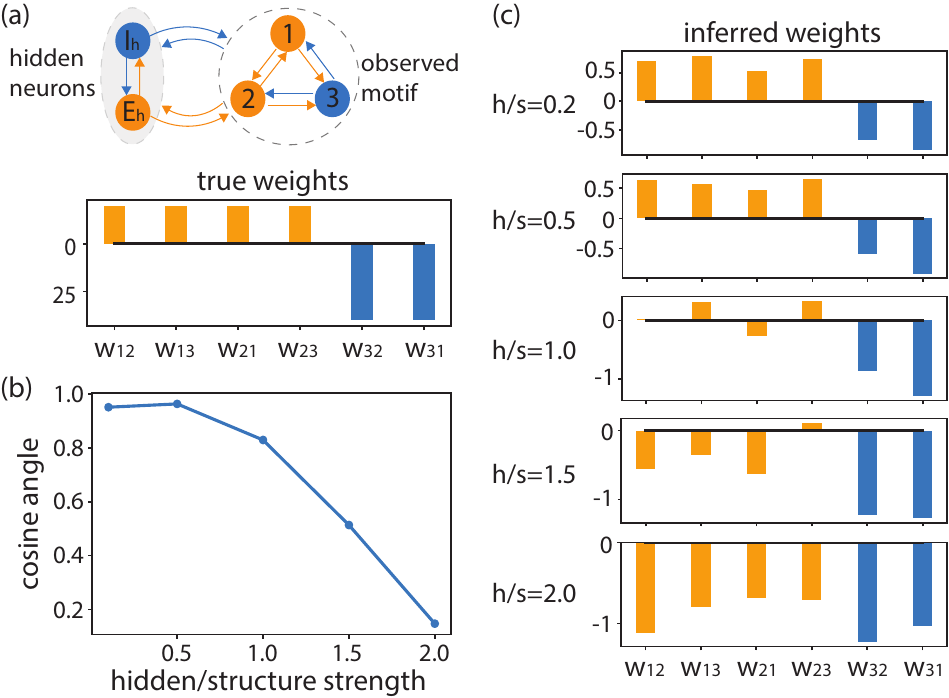}
\par\end{centering}
\caption{Effects of hidden neurons. \textbf{(a)} Schematic of a three-neuron motif that is observed and also connected with two hidden neurons. The true weights for two excitatory neurons and one inhibitory neuron are shown below. \textbf{(b)} The cosine angle between inferred and true weights as a function of the hidden-to-structure connected strength (h/s). \textbf{(c)}
 Inferred weights of the observed neural motif from weak (top) to strong (bottom) connections from hidden neurons (h/s). The quantified cosine angle is shown in (b).
\label{fig: C5_3}}
\end{figure}

We simulated a network of 5 LIF neurons, then only measure a sub-motif of three neurons to test how unobserved neural signals influence the inference performance (Fig.~\ref{fig: C5_3}a). The observed motif is the same E-I circuit analyzed in Fig.~\ref{fig: LIF} and the unobserved ones consist an excitatory and an inhibitory neuron. The unobserved and observed neurons are full connected. We scanned through different strength of interaction between the hidden and observed neurons and found that the inference is robust within a range of hidden neural input (Fig.~\ref{fig: C5_3}b,c).\\

We can correctly identify the synaptic weights and signs even when the hidden neurons are of the same order (half) of the strength of the observed network motif. This result indicates that one can confidently identify motifs from sub-sampled experimental measurements, under the mild assumption that the hidden neurons coupled to the observed units with slightly weaker weights. 



\begin{figure}[!t]
\begin{centering}
\includegraphics[width=1\columnwidth]{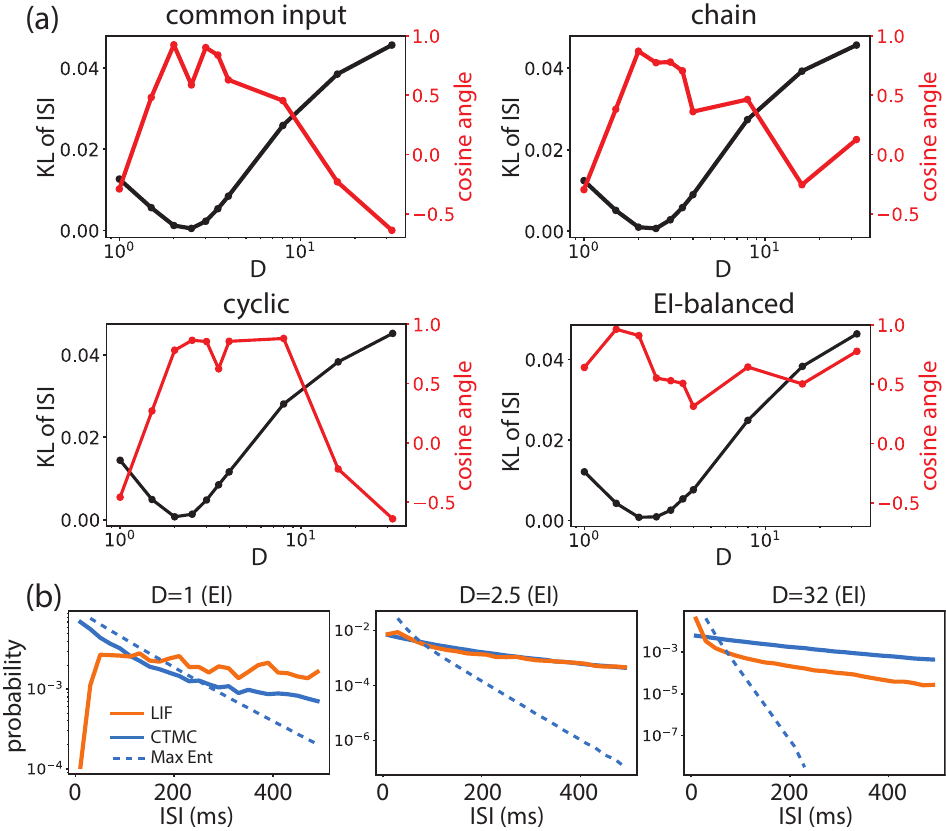}
\par\end{centering}
\caption{Spike train reconstruction and network inference. \textbf{(a)} The KL divergence between ISI of LIF simulation and CTMC reconstruction as a function of noise strength, $D$, is shown in black. The cosine angle between LIF network weights and inferred $w_{ij}$ is shown in red. The same analysis is done for four different network motifs. \textbf{(b)} Example ISI distribution of the E-I balanced circuit in (a), with three different noise strength $D$ selected. The empirical measurement from LIF spike train, CTMC reconstruction, and Max Ent prediction are shown.
\label{fig: ISI}}
\end{figure}

\subsection{Max Cal Reconstruction of Network Spike Train}

The Max Cal framework enables us to generate spike trains with the CTMC posterior model. We characterized the network spike train generated from CTMC with its inter-spike interval (ISI) distribution for the whole network (the sum of each neuron's ISI distribution).  
We explored how well CTMC recapitulates the ISI of LIF simulations across different aforementioned network motifs and input noise strength (Fig.~\ref{fig: ISI}). The Max Cal reconstruction is evaluated with the KL divergence between the ISI distribution measured from LIF spike trains and the CTMC reconstruction. We found consistent U-shape tuning to noise of this reconstruction across motifs, meaning that the network is best described by CTMC at a certain noise strength. Furthermore, the noise regime that generates spike trains well-approximated by CTMC coincides with better weight inference among different motifs (Fig.~\ref{fig: ISI}a). We also observed that coupling strength inference is more robust than ISI under varying noise levels across different motifs, showing that coupling strength inference is an task easier than ISI.\\

We studied the reconstruction of ISI distribution across three noise levels in the E-I circuit (Fig.~\ref{fig: ISI}b). To understand how much temporal structure is captured in the CTMC minimal model from Max Cal, we constructed a Max Ent model with discrete time bins for comparison \cite{schneidman_weak_2006}. We used discrete time windows to measure the long-term network-state distribution and after which we roll a $2^N$-face dice repeatedly to reconstruct spike trains from Max Ent. The resulting ``process'' would be independent and identically-distributed (i.i.d.) sampling of the measured distribution, with no temporal correlation. The ISI distributions inferred by Max Ent are depicted by dashed lines in Fig.~\ref{fig: ISI}b.\\

One sees clearly that the CTMC representation is a correction to the empirical ISI distribution generated by Max Ent and matches better to that of LIF circuits across all noise. Consistent with the U-shape KL divergence as a function of noise strength in Fig.~\ref{fig: ISI}a, we found that ISI distribution is best capture by CTMC at an intermediate noise level. Particularly, the CTMC misses the drop of distribution in low noise, and it over estimates long ISI in high noise cases. \\


\begin{figure}
\begin{centering}
\includegraphics[width=1\columnwidth]{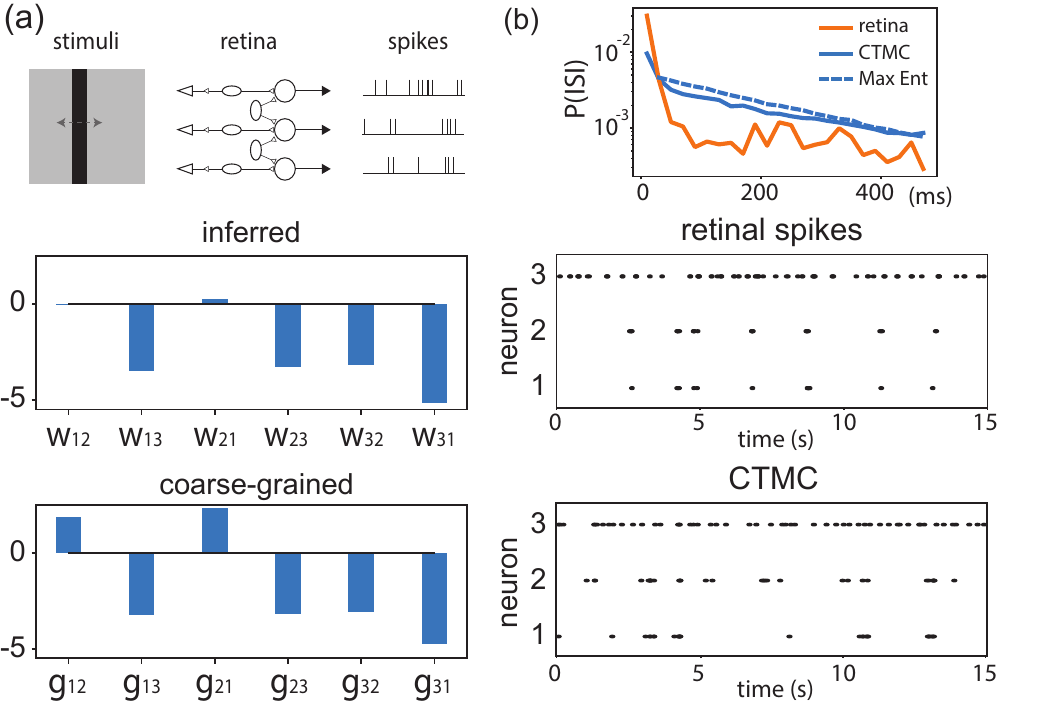}
\par\end{centering}
\caption{{Network inference with retinal spike trains.} \textbf{(a)} Schematic of the retinal spike trains measured under a stochastic moving-bar stimuli (top). The inferred weights $w_{ij}$ (middle) and coarse-grain coupling $g_{ij}$ (bottom) of three selected retinal ganglion cells are shown. The inferred $w_{12}=-0.05$ is weak and inconsistent with $g_{
12}$. \textbf{(b)} The ISI distribution (top) of retinal spike trains from the three neurons in (a). The measurement is compare to CTMC reconstruction and Max Ent. The measured (middle) and CTMC generated (bottom) network spike trains are shown.
\label{fig: retina}}
\end{figure}

\subsection{Application to Retina Data}

We explored network inference for retinal spike trains measured from Palmer \textit{et al.} \cite{palmer2015predictive}. The population spikes are measured from a group of salamander retinal ganglion cells driven by complex visual stimuli. We processed the spike train into network states with a 20 ms sliding window. In an example triplet of retinal ganglion cells, we found that a pair of neurons that may be mutually exciting and inhibiting the third one (Fig.~\ref{fig: retina}a). The result is consistent with previous results showing positive and negative effective coupling across retinal spike trains \cite{pillow2008spatio, schneidman_weak_2006}.\\

Similar to reconstruction of LIF spike trains in Fig.~\ref{fig: ISI}, we reconstructed the ISI distribution using the inferred CTMC representation and the Max Ent fit. We consistently found that CTMC is slightly closer to the empirical retinal spike train distribution. While the reconstructed spike train appear qualitatively similar (Fig.~\ref{fig: retina}b), however, the mismatch of ISI across intermediate time scale ($\sim$100 ms) is still prevalent. This is different from the mismatch trends in noise-driven cases in Fig.~\ref{fig: LIF}b. One hypothesis is that the retinal spike train is a result of both the dynamical visual stimuli and more intricate pre-synaptic computation. The idea is supported by the fact that we observed extra effective coupling structure beyond the time scale of single neurons (SFig.~\ref{SI:retina_window}). While future work is needed to incorporate more elaborate encoding model in the framework, the consistent weight inference validates that the method can be generalize to biological measurements.

\section{Discussion \label{sec: discussion}}

In this work, we applied Maximum Caliber (Max Cal) inference to the jump-process representation of network spike trains to recover network and response properties with the inferred Continuous-Time Markov Chain (CTMC) as a minimal model. For numerical simulations of LIF neurons, we can infer the synaptic couplings and neural responses. The inference is applicable across different neural motifs and generalizes to cases with unobserved neurons, from which we demonstrated that our method can extract causality, not correlation, and direct couplings instead of indirect effective couplings. The inter-spike interval (ISI) distribution can be reconstructed through Max Cal in a range of input noise strength to the network.
Lastly, we applied our method to spike trains measured from the retina. While the ISI reconstruction has a larger mismatch, potentially due to strong stimuli applied to the system, we were able to characterize the retinal motif with this method.\\



\paragraph{Sensitivity to finite data.}

The counting statistics of a Markov chain, such as the state occupancy time and transition counts, vary with finite measurements. This is particularly the case for spike trains in finite data simulation and experimental measurements. To explore the effects of finite data, we numerically computed learning curves of finite data and found that the trend are consistent with the infinite-data idealized case (SFig.~\ref{SI:finite_data}). 
To compare simulation with experimental measurements, we verified that the state transition counts in the retina data is comparable to those in the simulated spike train sufficient for network inference (order of $\sim$1000 transitions). Lastly, we confirmed that perturbing the time window size for the retinal spike trains has similar effect on weight inference as the simulated spikes (SFig.~\ref{SI:window},\ref{SI:retina_window}).
These results indicates that conclusion can be drawn from finite but sufficient measurements.\\

\paragraph{More neurons.}

The CTMC representation of spike trains grow exponentially with the number of observed neurons. This is a similar challenge for past work building state space for neural measurements \cite{lynn_decomposing_2022, roudi2009statistical}. While mainly studying various 3-neuron motifs in this work, we have shown that this 3-neuron spike trains could be a subset of the total network of 5 neurons, given that the ignored couplings are slightly weaker. 
According to the setup in Fig.~\ref{fig: rates}, the degree of freedom (dof.) for 5 neurons grows to 160, in comparison to 32 dof for 3 neurons. As dof grows with the size of the hypercube, the demand of data for sufficient statistics becomes large, and should be an important future direction. One possible alternative is to construct lower dimensional state-space that may not scale with the number of measured units.\\

An interesting benefit from considering larger network is the richer motifs and topology. As shown in previous work \cite{haber2022learning}, the possible motifs with directed connections is on the order of thousands for only 4 neurons. Further, following our network-state jump-process framework, one can study different cycles and flux across motifs and discuss the how ``entropy production'' is related to the motif functions \cite{lynn2021broken,lynn_decomposing_2022}. \\





\paragraph{Comparison to other inference methods}

We focused on small network motifs \cite{milo2002network, luo2021architectures} for starters to test our method, as couplings in small motifs ---with indirect/cyclic couplings and heterogeneous neurons--- could already be hard to infer \cite{prinz2004similar, gerhard2013successful, haber2022learning}. Prior work applied Max Ent models to characterize functional coupling that are related to correlation of the spike trains \cite{schneidman_weak_2006, roudi2009statistical}. In protein folding, it has been shown that Max Ent can distinguish between direct and indirect coupling \cite{morcos2011direct}, but the dynamics generalization to coupling with time delays is not known, which our present method paves the way to solve that.\\ 

The causal coupling has been explored in pair-wise interaction frameworks such as the generalized linear model (GLM) or Granger causality \cite{pillow2008spatio, ladenbauer2019inferring, gerhard2013successful}. Empirically, it has been shown that GLMs can successfully reconstruct network connectivity \cite{gerhard2013successful, das2020systematic, ladenbauer2019inferring}. However, common across all methods and our results, the inference is particularly challenging in strongly interacting networks and when units are unobserved \cite{das2020systematic, liang2024statistically}. Our framework is similar to GLM in characterizing causal interactions, but generalizes from pair-wise interaction to network states that include higher-order interactions. Our approach is similar to Max Ent in finding minimal models given data statistics, but Max Cal model offers a novel, principled approach to infer network properties and spiking patterns. The present paper aims to fully explore our framework as a new way to infer network properties. We will leave detailed methods performance comparison to future works.\\




\paragraph{Future directions}

While we analyzed CTMC from network spike trains, this method does not require the underlying neural dynamics to be Markovian. Our approach captures the first-order temporal interaction, and we show that many of these descriptions are consistent with the underlying biophysical parameters in simulation. Extending the current framework, one could consider more observables, such as state dwell time moments and distribution \cite{skinner_estimating_2021}, in Max Cal to improve the inference.  Moving beyond stationary spike trains, future work can also incorporate more complex dynamics in the parameters, such as short-term synaptic plasticity or learning in the coupling parameters \cite{gerstner_neuronal_2014, koch2004biophysics}. In addition, in this work we examine our framework mainly by noise-driven spontaneous spiking activity in a network. It would be interesting to consider how to extend the framework for networks driven by more coherent stimuli \cite{lynn_decomposing_2022, chen2020nonequilibrium, palmer2015predictive}.
Neural spike trains enables computation such as encoding readout \cite{pillow2008spatio, rieke1999spikes}. Another promising future direction is to study how these network states are used in information processing \cite{lynn2021broken, granot2013stimulus}.



\begin{acknowledgments}
The authors thank Rostam Razban for helpful feedback. YJY thanks the support of the Laufer Center. KSC thanks support from PNI, CPBF, and the Princeton University Library for funding open source submission.
\end{acknowledgments}

\appendix

\section{Max Cal Computation}

The Max Cal objective in Eq. \eqref{eq: Max Cal} was optimized with Sequential Least Squares Programming (SLSQP) in Python. The equality constrain imposed by minimizing the squared error between model and observables and the scalar objective is the KL term. We used a uniform stationary prior for $K^0_{ij}$. We empirically verified that the tolerance and initial conditions have negligible effects on the convergence. Upon convergence, the equality constraints are down to the order of $10^{-12}$. Data and code used to generate figures in this work are available: \url{https://github.com/Kevin-Sean-Chen/MaxCal_network}

\section{Spiking Neural Network Simulation}

We simulated leaky-integrate-and-fire (LIF) circuits using Euler method with time step $dt$ as 0.1 ms. The voltage dynamics of the $i$-th neuron follows: 
\begin{equation}
    \tau_m\frac{dv_i}{dt} = v_r - v_i + RI_i(t) + \xi_i
\end{equation}
where $v$ is rest to $v_r=-65$ mV if it passes threshold $v_{tr}=-50$ mV and marked as a spike. The membrane time constant $\tau_m=10$ ms and resistance $R=1$ are fixed for all neurons. The time-varying white noise $\xi$ is white noise with amplitude $D$, following $\langle \xi(t)\xi(t') \rangle = 2D\delta(t-t')$. The synaptic current $I$ follows:
\begin{equation}
    \frac{dI_i}{dt} = \frac{-I_i}{\tau_s} + \Sigma_j w_{ij} \Sigma_{t_{jk}<t} \delta(t-t_{jk})
\end{equation}
where $\tau_s=5$ ms is the synaptic time constant, $w_{ij}$ is the synaptic weights from neuron $j$ to neuron $i$, and the $k$th spike from neuron $j$ at time $t$ is denote $t_{jk}$. In practice, we simulated $\sim$100 seconds of spikes, then embedded the network spike train onto the network states.

\section{Effects of Window Size and Signal to Noise }

The size of the sliding window would affect state transition structure. We analyzed the window size with respect to the inter-spike interval distribution of LIF network. At small window size, the inferred weights are biased towards negative value. This is due to many observed single neuron states that transition back to full-silence and the lack of richer network states when the bin size it close to the refractory period. We empirically observed that the inferred weights are robust within a range of intermediate time window size (SFig.~\ref{SI:window}). Unless mentioned otherwise, we used window size $B=20$ms throughout the work, for both analysis of LIF models and retinal spike trains.\\

\begin{suppfigure}[!h]
\centering
\includegraphics[width=1.\columnwidth]{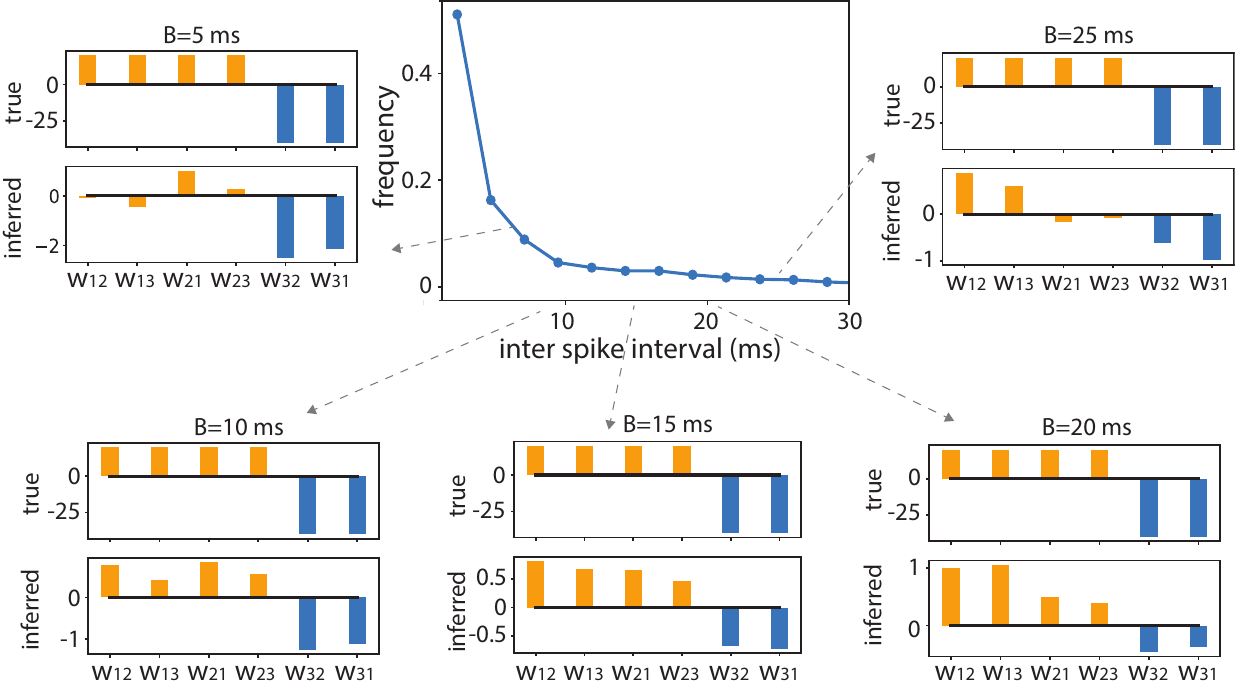}
\caption{Window size perturbative effects on inference. The inter-spike interval across three neurons are shown in the middle curve. We compute states using five different window sizes $B$ and infer the coupling weights $w_{ij}$. The results show that the conclusion for sign and trend of the weights are robust within a window range.
}
  \label{SI:window}
\end{suppfigure}

The signal from network structure is controlled by a scalar $\alpha$ that rescaling network by $\alpha w_{ij}$ and the noise amplitude $\langle \xi(t)\xi
(t') \rangle = 2D\delta(t-t')$. We scanned through a range of $\alpha$ and $D$ and record the inference performance for LIF network simulations. In agreement with the signal-to-noise ration, the correlation between inferred weights and true synaptic connections are large at strong network strength and weak noise strength (SFig\ref{SI:SNR}). Unless mentioned otherwise, we used $D=2$ and $\alpha=20$ throughout the LIF simulations.

\begin{suppfigure}[!h]
\centering
\includegraphics[width=.7\columnwidth]{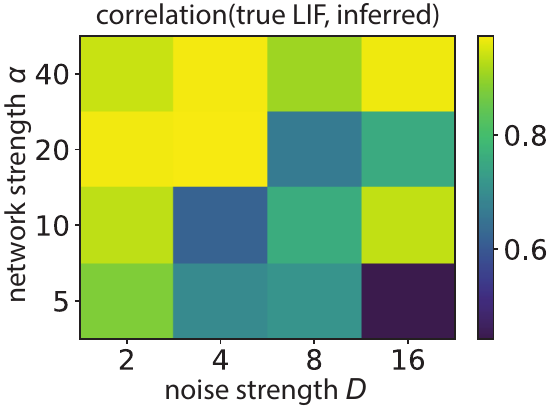}
\caption{The correlation coefficient between inferred weights and the true synaptic connection in LIF, for different network strength and noise strength.
}
  \label{SI:SNR}
\end
{suppfigure}

In the learning curve shown in Fig.~\ref{fig: dof}, the summary statistics for CTMC computes the analytic infinite time result. We sought to test for finite-data effects by simulating only 50 steps of the 3-neuron CTMC model. We found that the finite measurements have slight bias away from the infinite time solution, but consistently follow the same trend (SFig.~\ref{SI:finite_data}).

\begin{suppfigure}[!t]
\centering
\includegraphics[width=1.\columnwidth]{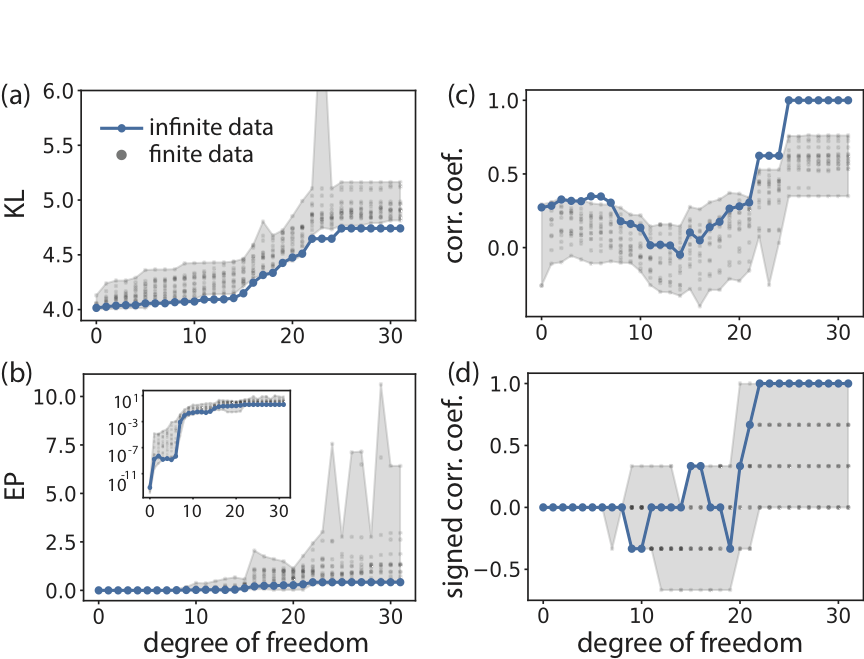}
\caption{Finite sample effects in learning curves and inference. \textbf{(a)} KL as a function of degree-of-freedom in CTMC sorted by the observed frequency. 15 simulated state trajectories are shown in grey dots, with area covering the minimum and maximum values from finite data for visual guidance. The result for infinite data (computed analytically) is shown in blue. \textbf{(b)} Same as (a) but for entropy production (EP) with log of EP shown in inset. \textbf{(c)} for correlation coefficient against the true parameters. \textbf{(d)} for signed-correlation of the coupling weights.
}
  \label{SI:finite_data}
\end
{suppfigure}

\section{Window Effect on the Inferred Retinal Coupling}

We demonstrated an example retinal neural motif with mutual excitation in Fig.~\ref{fig: retina}a. Empirically, however, most of the triplets we selected from measurements show negative coupling or only weak excitation (SFig.~\ref{SI:retina_window}a). This might be consistent with known physiology of the retinal ganglion cells (RGC) ---- there is unlikely to be directly synaptic connection across RGCs, and many RGCs have lateral inhibition between each other \cite{luo2021architectures, rieke1999spikes}. The inference reveals consistent negative weights at the 20 ms time scale, which is short enough for direct synaptic coupling.

The retinal spike train conveys information about the complex visual stimuli, in this case a moving bar following a random walk \cite{palmer2015predictive}. As indicated in the paper \cite{palmer2015predictive}, the stimulus has correlation on the time scale of $100-200$ms and moves continuously across the receptive filed of the measured RGCs. We sought to characterize such ``functional coupling'' induced by spatiotemporally correlated external input by re-analyzing spike train with a larger window (SFig.~\ref{SI:retina_window}b). In result, we found that at larger window size $B=150$ ms, the inference shows more diverse functional coupling with more excitatory coupling. We hypothesize that it reflects the long time-scale signal introduced through stimulus or more complex retinal computation, which can be studied through more complex encoding models in the future \cite{pillow2008spatio, gerstner_neuronal_2014, granot2013stimulus}.

\vfill\eject

\begin{suppfigure}[!h]
\centering
\includegraphics[width=1.\columnwidth]{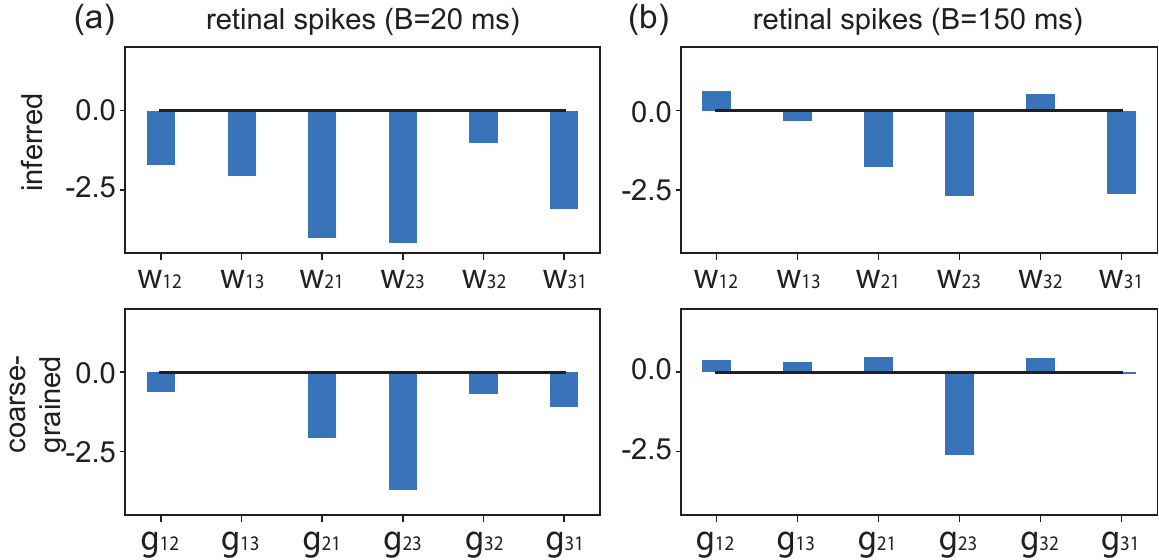}
\caption{Time window effect on coupling inferred from retinal spike trains. \textbf{(a)} Inferred coupling (top) and coarse-grained (bottom) analysis for a selected triplet of retinal ganglion cells. The sliding window size is $B=20$ ms. \textbf{(b)} Same triplet of spike trains analyzed with $B=150$ ms window size.
}
  \label{SI:retina_window}
\end
{suppfigure}

\bibliography{MaxCal-NN.bib}
\end{document}